\documentstyle[12pt]{article}
\def\b{\begin{equation}} \def\e{\end{equation}}
\def\bd{\begin{displaystyle}} \def\ed{\end{displaystyle}}
\def\ba{\begin{array}} \def\ea{\end{array}}

\def\bee{\begin{enumerate}}
\def\eee{\end{enumerate}}

\def\bes{\begin{eqnarray*}}
\def\ees{\end{eqnarray*}}
\def\be{\begin{eqnarray}}
\def\ee{\end{eqnarray}}

\def\1{\mbox{I\hspace{-.15em}1}}

\def\b{\begin{equation}}
\def\e{\end{equation}}
\def\bee{\begin{enumerate}}
\def\eee{\end{enumerate}}

\begin{document}

\title{Krein Regularization of $\lambda\phi^4$}

\author{ B. Forghan$^{1}$\thanks{e-mail:
b.forghan@piau.ac.ir}}

\maketitle  \centerline{\it $^1$ Department of Physics, Science
and Research Branch, } \centerline{\it Islamic Azad University,
Tehran, Iran}

\begin{abstract}

We calculate the four-point function in $\lambda\phi^4$ theory by
using Krein regularization and compare our result, which is
finite, with the usual result in $\lambda\phi^4$ theory. The
effective coupling constant  $(\lambda_{\mu})$ is also calculated
in this method.

\end{abstract}

\vspace{0.5cm} {\it Proposed PACS numbers}: 04.62.+v, 03.70+k,
11.10.Cd, 98.80.H \vspace{0.5cm}

\section{Introduction}

Due to the appearance of infrared divergence in the two point
function for the minimally coupled scalar field in de Sitter
space, a new method of field quantization has been presented
\cite{gareta1,dbre}. A covariant quantization of minimally coupled
scalar field has been constructed by positive and negative norm
states \cite{gareta1,thesis}. In the same manner as in the Gupta -
Blueler quantization of the electrodynamic equations in Minkowski
space, we have performed the field quantization in
 Krein space \cite{gup,bleu}.

It was conjectured that quantum metric fluctuations might smear
out the singularities of Green functions on the light cone, but
they do not remove other ultraviolet divergences of quantum field
theory \cite{for2}. However, it has been shown that quantization
in Krein space removes all ultraviolet divergences of QFT except
the light cone singularity \cite{gareta1,ta3,refaei,ba,rota}.

It has been shown that the Krein propagators which obey the Krein
quantization and quantum metric fluctuation are finite
\cite{rota,rerota,zft,ftz}. The most interesting result of this
construction is a new method of regularization, which we have
called "Krein regularization". A natural regularization of the
one-loop approximation for an interacting quantum scalar field in
Minkowski space $(\lambda\phi^4)$ has been achieved through the
application of Krein space quantization \cite{ta4}. The Casimir
effect and one-loop approximation of Moller scattering have been
calculated in Krein space \cite{knrt,pay}. In QED, the value of
the Lamb shift  was calculated in Krein space quantization
including quantum metric fluctuation and the  magnetic anomaly was
computed in Krein space quantization \cite{ftz}.

In this paper, we calculate the four-point function in Krein space
quantization including quantum metric fluctuation through the
application of Krein regularization. In this method, the effective
coupling constant is also calculated.

\setcounter{equation}{0}
\section{Scalar Green Function in Krein space}

In terms of  positive and negative norms, the field operator in
Krein space quantization can be written as \cite{gareta1}: \b
\phi(x)=\frac{{1}}{{\sqrt2}}[\phi_p(x)+\phi_n(x)],\e where $
\phi(x)$ satisfies the Klein - Gordon equation and $$
\phi_p(x)=\int
d^3\textbf{k}[a(\textbf{k})U_p(k,x)+a^\dag(\textbf{k})U_p^*(k,x)],$$\b
\phi_n(x)=\int
d^3\textbf{k}[b(\textbf{k})U_n(k,x)+b^\dag(\textbf{k})U_n^*(k,x)].\e
$a(\textbf{k})$ and $b(\textbf{k})$ are two independent operators.
Two sets of solutions are given by
$U_p(k,x)=\frac{{e^{-ikx}}}{{\sqrt{(2\pi)^32k_0}}}$ and
$U_n(k,x)=\frac{{e^{ikx}}}{{\sqrt{(2\pi)^32k_0}}},$ (with the sign
of the metric $( + , - , - , - )$) where
$k_0=\sqrt{\textbf{k}.\textbf{k}+m^2}\geq0$. Note that $U_n$ has
the negative norm.

The time-ordered product propagator for the scalar field is
defined as \cite{gareta1}: \b iG_T(x,x')=<0\mid T\phi(x)\phi(x')
\mid 0>=\theta (t-t'){\cal W}(x,x')+\theta (t'-t){\cal W}(x',x).\e
In this case we obtain: \b
G_T(x,x')=\frac{1}{2}[G_F(x,x')+(G_F(x,x'))^*]=\Re G_F(x,x'),\e
where the Feynman Green function is defined by \cite{bida}:

$$ G_F(x,x')=\int
\frac{d^4 k}{(2\pi)^4}e^{-ik.(x-x') }\tilde G_F(k)=\int \frac{d^4
k}{(2\pi)^4}\frac{e^{-ik.(x-x')}}{k^2-m^2+i\epsilon}=$$ \b
-\frac{1}{8\pi}\delta
(\sigma_0)+\frac{m^2}{8\pi}\theta(\sigma_0)\frac{J_1
(\sqrt{2m^2\sigma_0})-iN_1 (\sqrt{2m^2\sigma_0})}{\sqrt{2m^2
\sigma_0}}-\frac{im^2}{4\pi^2}\theta(-\sigma_0)\frac{K_1
(\sqrt{-2m^2\sigma_0})}{\sqrt{-2m^2 \sigma_0}},\e in which
$2\sigma_0=(x-x')^2=\eta_{\mu\nu}(x^\mu-x'^\mu)(x^\nu-x'^\nu)$.
Therefore: \b G_T(x,x')=\int \frac{d^4
k}{(2\pi)^4}e^{-ik.(x-x')}PP\frac{1}{k^2-m^2}=-\frac{1}{8\pi}\delta
(\sigma_0)+\frac{m^2}{8\pi}\theta(\sigma_0)\frac{J_1
(\sqrt{2m^2\sigma_0})}{\sqrt{2m^2 \sigma_0}}, \;\;x\neq x'.\e

The contribution of the coincident point singularity $(x=x')$
merely appears in the imaginary part of $G_F(x,x')$  (\cite{ta3}
and equation (9.52) in \cite{bida})
$$G_F(x,x')=-\frac{2i}{(4\pi)^2}\frac{m^2}{d-4}+G^{finite}(x,x'),$$
where d is the space-time dimension and $G^{finite}(x,x')$ becomes finite as $d\rightarrow4$.

In the momentum space for this propagator we have \cite{itzu} \b
\tilde G_T(k)=\frac{1}{2}[\tilde G_F(k)+\tilde
G_F(k)^*]=\frac{1}{2}\left[\frac{1}{k^2-m^2+i\epsilon}+
\frac{1}{k^2-m^2-i\epsilon}\right]=PP \frac{1}{k^2-m^2}.\e

The quantum field theory in Krein space, including the quantum
metric fluctuation $(g_{\mu\nu}=\eta_{\mu\nu}+h_{\mu\nu})$,
removes all ultraviolet divergencies of the theory
\cite{for2,rota}: \b \langle G_T(x - x')\rangle = -\frac{1 }{8\pi}
\sqrt{\frac{\pi}{2\langle\sigma_1^2\rangle}}
exp\left(-\frac{\sigma_0^2}{2\langle\sigma_1^2\rangle}\right)+
\frac{m^2}{8\pi}\theta(\sigma_0)\frac{J_1(\sqrt {2m^2
\sigma_0})}{\sqrt {2m^2 \sigma_0}},\e where
$2\sigma=g_{\mu\nu}(x^\mu-x'^\mu)(x^\nu-x'^\nu)$ and
$\sigma=\sigma_0+\sigma_1+O(h^2)$. The average value is taken over
the quantum metric fluctuations.

The Fourier transformation of the second part of the equation
$(2.8)$ is \cite{rerota,ta4}:\b  PP\frac{m^2}{k^2(k^2-m^2)}.\e
This propagator has been used by some authors to improve the UV
behavior in relativistic higher-derivative correction theories
\cite{ba,ho} and also appears in the supersymmetry theory
\cite{kaku}.

For diagrams with loops there are two possibilities: the diagram
is convergent or divergent due to the singularity of the delta
function. In the first case we chose the propagator as
$PP\frac{1}{k^2-m^2}$. For the second case we selected the
propagator as $(2.9)$ which would result in the answer being
finite and with no need for any renormalization. This means there
are no added  counter terms in the Lagrangian.

\setcounter{equation}{0}
\section{$\lambda \phi^4$ in Krein Space Quantization}

The S matrix elements are the most important quantities in the
interacting QFT, which can be written in terms of the time order
product of the two free field operator by applying the LSZ
reduction formulas, Wick's theorem and time evolution operator
\cite{itzu}.

The un-physical states  can not propagate in the physical system
because they are not observed in nature, so the following
conditions have been imposed on the physical states for
eliminating these unphysical states:
$$ b(k)|\mbox{physical states}\rangle=0.$$

The tree order S-matrix elements do not change because the
propagator in the two methods is the same $(k^2\neq m^2)$
\cite{zft}.

In the $\lambda\phi^4$ theory, for calculating the four - point
function the negative norm states appear in the calculations of
the loop expansion, where they cause the negative mode states to
propagate in the loop.

\subsection{Four-Point Green Function}

By using the S-matrices elements \cite{huang}: $$ \mathop
s\nolimits_{fi} =\left\langle {in, p_3,p_4}
\right|1+S^{(1)}+S^{(2)}+...\left| {p_1 ,p_2 ,in }
\right\rangle.$$  In studying the second-order Feynman graphs only
$S^{(2)}$ is important which is:
 \b S^{(2)}  = \frac{1} {{2!}}\left( { -
\frac{{i\lambda }} {{4!}}} \right)^2 \int {d^4 x_1 } \int {d^4 x_2
} T\left[ {\varphi ^4 (x_1 )\varphi ^4 (x_2 )} \right].\e  There
are 3 different diagrams: the $s,t$ and $u$ channels. The
variables $s,t$ and $u$ are known as mandelstam variables.
$\Gamma^{(4)}(t)$ and $\Gamma^{(4)}(u)$ are obtained from
$\Gamma^{(4)}(s)$ by interchanging $p_{3}$ and $p_{4}$, $p_{2}$
and $-p_{3}$  and the sum of these three graphs are:
$$\left\langle {p_3,p_4 } \right|S^{(2)} \left| {p_1 ,p_2}
\right\rangle =\Gamma^{(4)}(s)+\Gamma^{(4)}(t)+\Gamma^{(4)}(u)=
$$\b \frac{{(-i\lambda)^2}}{{2}}[I(p_1,p_2,p_3,p_4)+I(p_1,p_2,p_4,p_3)+I(p_1,-p_3,-p_2,p_4)]K_{1234},\e
where $K_{1234}  = \frac{{(2\pi )^4 \delta ^4 (p_1  + p_2  - p_3 -
p_4 )}} {{\sqrt {2\omega _1 2\omega _2 2\omega _3 2\omega _4 }
}},\omega _i  = \sqrt {p_i^2  + m_0^2 } $ and

\b I(p_1,p_2,p_3,p_4) = \int {\frac{{d^4 k}} {{(2\pi )^4 }}}
\widetilde{G}_T (k)\widetilde{G} _T (k + p),\e

By including the quantum metric fluctuation, $\widetilde{G}_T (k)$
and $\widetilde{G} _T (k + p)$ in the above equation must be
replaced by $<\widetilde{G} _T (k)>$ and $<\widetilde{G} _T (k +
p)>$.  Since $<\widetilde{G} _T (k)>$ and $<\widetilde{G} _T (k +
p)>$ are finite, the four-point function in this formalism is
automatically regularized and no divergent term is encountered

\b \Gamma^{(4)}_{kr}(p) = \frac{3\lambda^2 } {2}\int {\frac{{d^4
k}} {{(2\pi )^4 }}} <\widetilde{G} _T (k)><\widetilde{G} _T (k +
p)>.\e

\setcounter{equation}{0}
\section{Krein Regularization for the Four-Point Function}

Because of the delta function singularity in the propagators, the
integral $(3.3)$ is divergent at the ultraviolet limit, whereas
the integral $(3.4)$ is finite:

\b \Gamma^{(4)} _{kr} (p)= \frac{3\lambda^2 } {8}\int \frac{d^4
k}{(2\pi )^4 } PP\left(\frac{1} {{k^2 - m^2 }} - \frac{1} {{k^2
}}\right)PP\left(\frac{1} {{(p - k)^2-m^2 }}-\frac{1} {{(p - k)^2
}}\right). \e

In order to solve this integral, we use Feynman parameters: $$
\Gamma^{(4)} _{kr} (p) = \frac{{3\lambda^2}}{{2}} \int\limits_0^1
{dx} \int {\frac{{d^4 l}} {{(2\pi )^4 }}} \left[\frac{1} {{(l^2
+x(1-x)p^2-m^2 )^2 }}\right. - $$\b \left. \frac{1} {{(l^2
+x(1-x)p^2-m^2(1-x))^2 }}-\frac{1} {{(l^2 +x(1-x)p^2-xm^2)^2
}}+\frac{1} {{(l^2 +x(1-x)p^2)^2 }}\right], \e where $ k=l - xp$
 \cite{chengli,pesc}.

The integral over $(l)$ is no longer divergent. In order to solve
the integral over $l$, we apply the Wick rotation  in a manner
that if $i\epsilon$ exists in the denominator, the substitution
$l_E^0 = - il^0$ is used, and if, conversely, $-i\epsilon$ is
present in the denominator, the change of variable $l_E^0 = il^0$
is applied and we arrive at the following integral \cite{pesc}:
 \b
\Gamma^{(4)}_{kr} (p) = -\frac{{3\lambda^2 }}{{32\pi
^2}}\int\limits_0^1 {dx} \left\{\ln
\left(1-x(1-x)\frac{{p^2}}{{m^2}}\right)\right.+\left.\ln
\left(\frac{{-p^2}}{{m^2}}\right)-\ln
\left(1-x\frac{{p^2}}{{m^2}}\right)-\ln
\left(1-(1-x)\frac{{p^2}}{{m^2}}\right)\right\} .\e

In Hilbert space, we have \cite{chengli}:
 \b\Gamma^{(4)}_{Hi} (p) = \frac{{3\lambda^2 }}{{32\pi
^2}}\int\limits_0^1 {dx} \left\{\frac{{2}}{{4-d}}-\ln
\left(1-x(1-x)\frac{{p^2}}{{m^2}}\right)\right\}.\e

\subsection{Effective Coupling Constant}
The effective potential was calculated in Krein space quantization
\cite{rerota}. It was used to calculate the $\beta$-function in
Krein space and the answer was:$$\beta = \frac{{d\lambda _{eff} }}
{{dt}} = \frac{{3\lambda ^2 }} {{16\pi ^2 }},$$ where
$\lambda_{eff}= \left. {\frac{{d^4 V_{eff} }} {{d\varphi ^4 }}}
\right|_{\varphi ^2  = \mu ^2 } $ , $\mu=e^{-t}$ and the
$\beta$-function was similar to it's counterpart in  Hilbert
space.

In our method, by defining the scattering matrix elements and
using equation $(4.3)$ at the scale of energy $p^2=-\mu^2$, the
effective coupling constant would be as below: \b \lambda _\mu
 = \lambda  + \frac{{3\lambda ^2 }} {{32\pi ^2
}}\int\limits_0^1 {dx} \left\{ {\ln \left( {1 + (x - x^2
)\frac{{\mu ^2 }} {{m^2 }}} \right) + \ln \left( {\frac{{\mu ^2 }}
{{m^2 }}} \right) - 2\ln \left( {1 + x\frac{{\mu ^2 }} {{m^2 }}}
\right)} \right\},\e which is finite. The $\beta$-function would
be defined as follows: \b \beta  = \mu\frac{{d\lambda _{\mu} }}
{{d\mu}} = \frac{{3\lambda ^2 }} {{16\pi ^2 }},\e which is in
agreement with the result from the previous methods.

\section{Conclusion}

In this paper we have explicitly calculated  the four-point
function in the one loop approximation. In Krein space
quantization including quantum metric fluctuation, the Green
function is finite and does not have any divergent term in the
ultraviolet and infrared limit.

Using the four-point function, the coupling constant is written at
the scale of  energy $\mu$ and the $\beta$-function is calculated.
 The result is similar to that gained from the effective potential and
Hilbert space methods.

In this method $\lambda \phi^4$ is automatically regularized and
renormalization is not applied. This method can be easily
generalized to non-Abelian gauge theory and quantum gravity in the
background field method. It can also be employed in the
calculation of the $\beta$-function in QED.

\vspace{0.5cm} \noindent {\bf{Acknowlegements}}: The author is
grateful to M.V.Takook for his helpful discussions.


\begin{thebibliography}{a}



\bibitem{gareta1} J.P. Gazeau, J. Renaud, M.V. Takook, Class. Quantum
Grav. $17(2000)1415$, gr-qc/$9904023$.
\bibitem{dbre}  S. De Bievre, J. Renaud, Phys. Rev. D $57(1998)
6230$.
\bibitem{thesis} M. V. Takook, "Th\'{e}orie Quantique des Champs Pour des Syst\`{e}ms
\'{E}l\'{e}menteries Massifs et a Masse Nulle sur l'eSpace - Temps
de Sitter", "Th\`{e}se de l' universite' Paris VII, $(1997)$.
\bibitem{gup} O.S.N. Gupta, Proc. Roy. Soc. A $63(1950)681$.
\bibitem{bleu} K. Bleuler, Helv. Phys. Acta $23(1950)567$.
\bibitem{for2} H.L. Ford, Quantum Field Theory in Curved Spacetime,
gr-qc/$9707062$.
\bibitem{ta3} M.V. Takook, Mod. Phys. Lett. A  $16(2001)1691$,
gr-qc/$0005020$.
\bibitem{refaei} A. Refaei, M. V. Takook, Phys. Lett. B $704(2011)326$.
\bibitem{ba} N.H. Barth , S.M. Christensen, Phys. Rev. D $ 28(1983)1876$.
\bibitem{rota} S. Rouhani , M.V. Takook , Int. J. Theor. Phys. $48(2009)2740$.
\bibitem{rerota}  A. Refaei, M.V. Takook , Mod. Phys. Lett. A $26(2011)31$.
\bibitem{zft} A. Zarei, B. Forghan, M.V. Takook, Int. J. Theor. Phys, $50(2011)2466$.
\bibitem{ftz} B. Forghan, M.V. Takook, A. Zarei,"Krein
Regularization of QED", in preparation.
\bibitem{ta4} M.V. Takook, Int. J. Mod. Phys. E  $11(2002)509
$, gr-qc/$0006019$.
\bibitem{knrt} H. Khosravi, M. Naseri, S. Rouhani, M. V. Takook,
Phys. Lett. B $640(2006) 48$, gr-qc/$0604036$.
\bibitem{pay} F. Payandeh, M.V. Takook, M. Mehrafarine, Sci. China. Ser. G-Phys. Mech. Astron.
 $52(2009)212$.
\bibitem{bida} N.D. Birrell, P.C.W. Davies, Cambridge University Press,
{\it QUANTUM FIELD IN CURVED SPACE} $(1982)$.
\bibitem{itzu} C. Itzykson, J-B. Zuber, McGraw-Hill, Inc. {\it Quantum Field
Theory} $(1988)$ .
\bibitem{ho} P. Horva , Phys. Rev. D  $79(2009)084008,$ arXiv: $0901.3775$.
\bibitem{kaku} M. Kaku, Oxford University Press, Quantum Field Theory $(1993)$.
\bibitem{huang} K. Huang, Wiley-Interscience , Quantum Field Theory
$(1998)$.
\bibitem{chengli} T. P. Cheng, L. F. Li, Clarendon Press, Gauge Theory of
Elementary Particle Physics $(2000)$.
\bibitem{pesc} E. Peskin, D.V. Schroeder, Perseus Books, An Introduction  in Quantum Field Theory $(1995)$.





\end{thebibliography}
\end{document}